Transport Research Arena (TRA) Conference

# Novel pre-emptive control solutions for V2X connected electric vehicles

Kai Man So[a], Gaetano Tavolo[a], Davide Tavernini[a], Marco Grosso[b], Sergio Pozzato[b], Pietro Perlo[b], Aldo Sorniotti[a]*

*[a]University of Surrey, Guildford, GU2 7XH, United Kingdom*
*[b]Interactive Fully Electrical Vehicles srl, La Loggia (TO), 10040, Italy*

**Abstract**

V2X technologies will become widespread in the next generation of passenger cars, and enable the development of novel vehicle control functionalities. Although a wide literature describes the energy efficiency benefits of V2X connectivity, e.g., in terms of vehicle speed profiling and platooning, there is a gap in the analysis of the potential of vehicle connectivity in enhancing the performance of active safety control systems. To highlight the impact vehicle connectivity could have on future active safety systems, this paper presents two novel control functions for connected vehicles, benefitting from the precise knowledge of the expected path and tire-road friction conditions ahead, as well as the current position of the ego vehicle. These functions, developed within recent and ongoing European projects, are: i) pre-emptive traction control; and ii) pre-emptive braking control.

## 1. Introduction

Powertrain electrification, together with vehicle connectivity to other road users, the infrastructure, and the cloud (referred to as V2X), are key features of next generation vehicles. A rather wide literature describes the energy efficiency benefits of V2X connectivity for functionalities such as platooning. However, there is a gap in the analysis of V2X for enhancing the performance of active safety controllers. According to Montanaro et al. (2019), vehicles can be used as moving sensors, providing data for cooperative tire-road friction estimation, intersection management, lane changing, etc. For example, the cloud could elaborate the information from several connected vehicles and determine the position of possible low tire-road friction patches, which would be transmitted to the approaching vehicles. Therefore, preview-based active safety controllers could take advantage of such V2X information.

---

* Corresponding author. Tel.: +44 (0)1483689688
  E-mail address: a.sorniotti@surrey.ac.uk

This paper presents two active safety controllers with road preview, developed within the European Horizon 2020 projects STEVE, TELL and Multi-Moby, which use implicit nonlinear model predictive control (NMPC) technology, based on the on-line solution of an optimization problem along a finite horizon. This is enabled by the progressive enhancement of the available real-time control hardware, and the introduction of computationally efficient solvers for nonlinear optimal control problems (Houska et al., 2011).

The first controller is a preview-augmented (or pre-emptive) traction controller, which embeds consideration of the predicted tire-road friction coefficient profile ahead, while considering the non-linearities of the tires and vehicle. In Scamarcio et al. (2022), preliminary simulations and experiments were performed to demonstrate that an NMPC-based traction controller with tire-road friction preview can pre-emptively reduce the wheel slip peaks and oscillations in acceleration maneuvers with abrupt tire-road friction coefficient reductions. This paper expands upon Scamarcio et al. (2022), with: i) further proof-of-concept experiments on an electric vehicle (EV) prototype; and ii) a simulation-based sensitivity analysis on the performance benefit of the proposed preview-based traction controller for vehicles with different values of the pure time delay of the powertrain in response to a torque request variation.

The second controller is a pre-emptive braking controller. This is an evolution of the automated trail braking controller in Zarkadis et al. (2018), which slows down the vehicle if its current speed is deemed safety critical with respect to the present reference yaw rate and tire-road friction coefficient. However, the formulation in Zarkadis et al. (2018) is only reactive, i.e., it limits vehicle speed when this already exceeds the value corresponding to the desired trajectory curvature and current friction limits, and therefore its interventions may occur too late to achieve stabilization. Since navigation maps and vehicle localization are commonly used in modern production passenger cars, the curvature of the path ahead can be considered approximately known, while next-generation V2X technologies can provide the future tire-road friction profile. Therefore, this paper presents a pre-emptive braking controller, which slows the vehicle in advance if the vehicle speed is deemed too high for the path curvature and friction level ahead. As no direct yaw moment is involved, the actuation only reduces the traction torque and/or generates a braking torque, i.e., it only involves longitudinal vehicle dynamics control. The proposed algorithm could be easily implemented in any modern vehicle layout, without interference with the operation of conventional stability controllers. The benefits of the controller are demonstrated through: i) a simulation study, showing the sideslip angle control capability of the new function along obstacle avoidance tests; and ii) a proof-of-concept real-time implementation on an automated EV prototype.

## 2. Experimental and simulation set-ups

The TELL/Multi-Moby EV prototype (Fig. 1(a)), with a front centralized on-board electric powertrain, is used as a case study for the pre-emptive traction controller. The vehicle for the assessment of the pre-emptive braking controller is the Zero Emission test Bed for Research on Autonomous driving (ZEBRA) of the University of Surrey (Fig. 1(b)), which is a modified Renault Twizy, i.e., an L7e two-seater electric quadricycle with a rear central electric motor. In both vehicles, the motor is connected to the wheels through a single-speed transmission, open differential, half-shafts, and constant velocity joints.

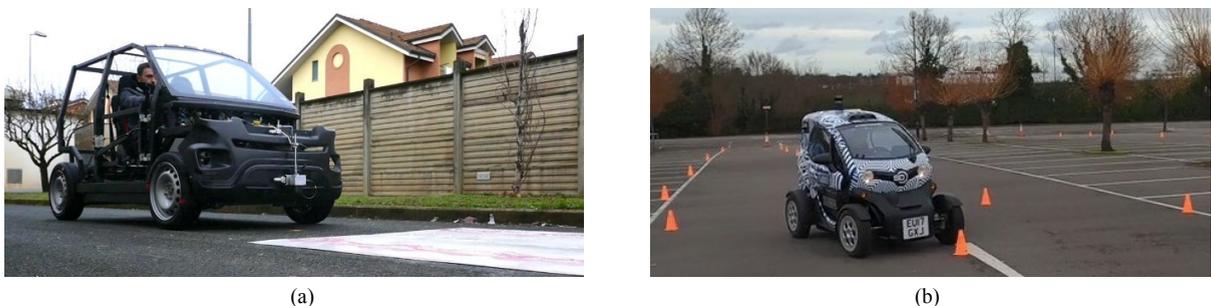

(a) (b)

Fig. 1. (a) TELL/Multi-Moby EV prototype during a traction control test with a step change from high (dry tarmac) to low (white boards covered with water and soap) tire-road friction coefficient; and (b) ZEBRA EV during a U-turn test (passive configuration) in the car park of the University of Surrey, UK.

Both EVs are equipped with i) individual wheel speed sensors; ii) a global positioning system (GPS); iii) an inertial measurement unit (IMU); iv) a Kistler sensor to optically measure the sideslip angle, and the lateral and longitudinal velocity components; and v) a dSPACE MicroAutoBox II system for rapid control prototyping.

Fig. 2 shows the simplified schematics of the control architectures, where the NMPC algorithms modify the torque level requested by the human or automated driver, to ensure either appropriate wheel slip levels (Fig. 2(a)) or a safe speed with respect to the upcoming trajectory (Fig. 2(b)). The controllers are implemented in Matlab-Simulink through the ACADO toolkit (Houska et al., 2011). The simulations are based on experimentally validated models of both vehicles in a Matlab-Simulink / IPG CarMaker environment, which is interfaced with the controllers. The experimental set-ups use the dSPACE system for the real-time solution of the nonlinear optimal control problem.

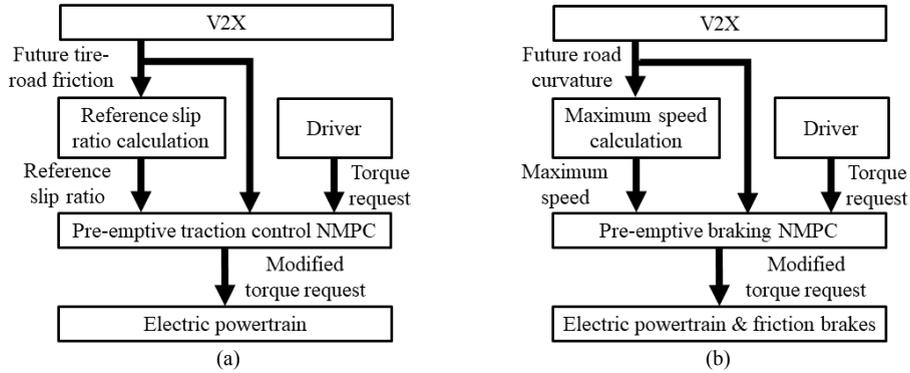

Fig. 2. Simplified schematics of the pre-emptive NMPC algorithms, based on V2X data, modifying the torque request from the human or automated driver for (a) traction control; and (b) automated braking to ensure safe cornering response.

## 3. Pre-emptive traction control

Two traction controllers were developed, i.e., the novel pre-emptive NMPC, and a benchmarking non-pre-emptive NMPC (i.e., a reactive formulation). The pre-emptive controller uses the predicted tire-road friction information in the context of V2X, while the control action of the non-pre-emptive controller is only based on the current tire-road friction condition, which can be obtained without V2X, e.g., from on-board state estimators.

### 3.1. Controller formulation

The traction controllers are defined for an EV with a front centralized on-board electric powertrain, but can be easily modified for other configurations. The internal NMPC model, used for the prediction, is expressed through the following continuous time formulation:

$$\dot{x}(t) = f(x(t), u(t)) \quad (1)$$

where the state vector $x$ is:

$$x = [\tau_m, s_{FL}, s_{FR}, \omega_{FL}, \omega_{FR}] \quad (2)$$

where $\tau_m$ is the motor torque; $s_{Fj}$ are the longitudinal wheel slip speeds, with the subscript $F$ indicating the front axle, and the subscript $j = L, R$ indicating the left or right sides of the EV; and $\omega_{Fj}$ are the angular wheel speeds.

The control action is defined as:

$$u = [\tau_{m,mod}, \varepsilon_{\sigma_x,FL}, \varepsilon_{\sigma_x,FR}] \quad (3)$$

where $\tau_{m,mod}$ is the motor torque demand after the modification by the traction controller; and $\varepsilon_{\sigma_x,Fj}$ are slack variables on the longitudinal tire slip ratios, which enable the implementation of a soft constraint. For conciseness, the full internal model equations are omitted from this paper; however, they are reported in Tavolo et al. (2022).

The nonlinear optimal control problem can be defined as:

$$\min_{u} J(x(0), u(\cdot)) := \sum_{n=0}^{N-1} l(x_n, u_n) \quad (4)$$

$$\text{s.t.}$$
$$x_0 = x_{in}(k)$$
$$x_{n+1} = f_d(x_n, u_n)$$

$$\underline{x} \leq x_n \leq \overline{x}$$
$$\underline{x} \leq x_N \leq \overline{x}$$
$$\underline{u} \leq u_n \leq \overline{u}$$
$$u(\cdot) : [0, N-1]$$

where $J$ is the cost function; $\boldsymbol{u}(\cdot)$ indicates the control sequence; $\boldsymbol{x}_{in}$ is the initial value of the state vector at the current time step $k$, obtained from the available sensor measurements and state estimators; $N$ defines the number of steps in the prediction horizon $H_P = N T_s$ with a constant time step $T_s$; $\underline{\boldsymbol{x}}$ and $\overline{\boldsymbol{x}}$ are the lower and upper limits for $\boldsymbol{x}$; $\underline{\boldsymbol{u}}$ and $\overline{\boldsymbol{u}}$ are the lower and upper limits for $\boldsymbol{u}$; $\boldsymbol{x}_{n+1} = f_d(\boldsymbol{x}_n, \boldsymbol{u}_n)$ is the discretized version of (1); and $l(\boldsymbol{x}_n, \boldsymbol{u}_n)$ is the stage cost, which is defined as:

$$l(\boldsymbol{x}_n, \boldsymbol{u}_n) = W_{u,\varepsilon_{\sigma_{x,FL}}} \varepsilon_{\sigma_{x,FL},n}^2 + W_{u,\varepsilon_{\sigma_{x,FR}}} \varepsilon_{\sigma_{x,FR},n}^2 + W_{u,\tau_m}[\tau_{m,driver} - \tau_{m,mod,n}]^2 \quad (5)$$

where $\tau_{m,driver}$ is the driver motor torque request in traction conditions; $W_{u,\varepsilon_{\sigma_{x,Fj}}}$ and $W_{u,\tau_m}$ are the cost function weights for the slack variables and control action penalization; and $n$ indicates the position of the step along the prediction horizon.

The constraints are expressed as:

$$0 \leq \tau_{m,mod,n} \leq \tau_{m,driver}$$
$$e_{Fj,n} + \varepsilon_{\sigma_{x,Fj},n} \geq 0 \quad (6)$$
$$\varepsilon_{\sigma_{x,Fj},n} \geq 0$$

The first line is a hard constraint, which states that the modified torque request in Fig. 2(a) can only be a reduction of the driver torque request, and not an increase. The remaining lines refer to a soft constraint on the longitudinal wheel slip ratio error $e_{Fj}$:

$$e_{Fj} = \sigma_{x,ref,Fj} - \sigma_{x,Fj} = \sigma_{x,ref,Fj} - \frac{s_{Fj}}{\omega_{Fj}R} \quad (7)$$

where $R$ is the wheel radius; $\sigma_{x,Fj}$ is the actual tire slip ratio; and $\sigma_{x,ref,Fj}$ is the reference tire slip ratio, which is expressed as a function of the estimated vertical tire load $F_{z,Fj}$ and tire-road friction condition $\mu_{Fj,fut}$:

$$\sigma_{x,ref,Fj} = f_{\sigma_{x,ref,F}}(\mu_{Fj,fut}, F_{z,Fj}) \quad (8)$$

The pre-emptive capability of the controller derives from $\boldsymbol{\mu}_{Fj,fut}$ being a vector (in this manuscript, vectors are indicated in bold), obtained from V2X, of the future tire-road friction condition values, computed from the vector $\boldsymbol{S}_{fut}(k)$ of the expected future traveled distance values along the prediction horizon, at the current time step $k$:

$$\boldsymbol{\mu}_{Fj,fut}(k) = f_{\mu_{Fj,fut}}(\boldsymbol{S}_{fut}(k) + \Delta x_{delay}(k)\mathbf{1}) \quad (9)$$

where in the implementation of this paper the function $f_{\mu_{Fj,fut}}$ is set as a map. For computational efficiency, $\boldsymbol{S}_{fut}(k)$ is generated under the constant speed ($V(k)$) assumption in the look-ahead period:

$$\boldsymbol{S}_{fut}(k) = S(k)\mathbf{1} + V(k)[\boldsymbol{t}_{fut} - t_{fut,0}\mathbf{1}] \quad (10)$$

where $S(k)$ is the current vehicle position; and $\mathbf{1}$ is an all-ones vector with dimension $N+1$. $\boldsymbol{t}_{fut} = [t_{fut,0}, t_{fut,1}, \ldots, t_{fut,N}]$ is the vector of future time values, defined for $N$ points evenly spaced according to the constant time step $T_s$, where $t_{fut,0}$ is the current time instant.

A novel feature of the pre-emptive controller is the delay compensation algorithm to account for the pure time delays between the driver torque request and the powertrain response, which can exceed 100 ms in typical EV implementations (Scamarcio et al., 2022). The delay compensator advances the map of the tire-road friction coefficient by a distance $\Delta x_{delay}$, corresponding to the pure time delay of the powertrain system, $\Delta t_{delay}$, under a constant speed assumption:

$$\Delta x_{delay}(k) = V(k)\Delta t_{delay} \quad (11)$$

In contrast, for the non-pre-emptive NMPC implementation, $\boldsymbol{\mu}_{Fj,fut}$ is a vector of identical components, and $\Delta x_{delay} = 0$. In both controllers, on top of being used for the reference slip ratio, $\boldsymbol{\mu}_{Fj,fut}$ is also provided to the tire model embedded in the prediction model.

### 3.2. Real-time controller implementation and experiments

Proof-of-concept experiments were performed, in which the driver requests full acceleration while the EV encounters a sudden transition from high to low friction conditions (Fig. 1(a)). The adopted settings are $H_P = 250$ ms and $N = 10$ steps (therefore $T_s = 25$ ms), which is feasible for real-time implementation. In the specific tests, the tire-road friction map was programmed a priori, while the current vehicle position was identified through vehicle speed integration due to the short test distance (while GPS is used for position identification in Scamarcio et al. (2022)).

Fig. 3 shows that the pre-emptive traction controller ('Pre-NMPC' in the figure) significantly improves the wheel slip control performance compared to the benchmarking NMPC without preview ('NMPC'), and the passive case. Subplot (a) includes the torque profiles, with a notable reference torque reduction for the Pre-NMPC (in blue) before the front driving wheels reach the low friction surface. A non-negligible pure time delay is observed between the requested and actual torque values, which confirms the importance of the proposed time delay compensator. Fig. 3(b) highlights the difference between the longitudinal vehicle speeds and the tangential wheel speeds computed from the angular wheel speeds. A large difference indicates significant wheel spinning, which is especially evident for the passive vehicle, travelling at only 10 km/h, but with wheels spinning at almost 80 km/h. Fig. 3(c) reports the wheel slip ratios, with the Pre-NMPC able to maintain a wheel slip ratio of <0.05, while the other configurations suffer from high slip ratios, with peaks exceeding 0.6.

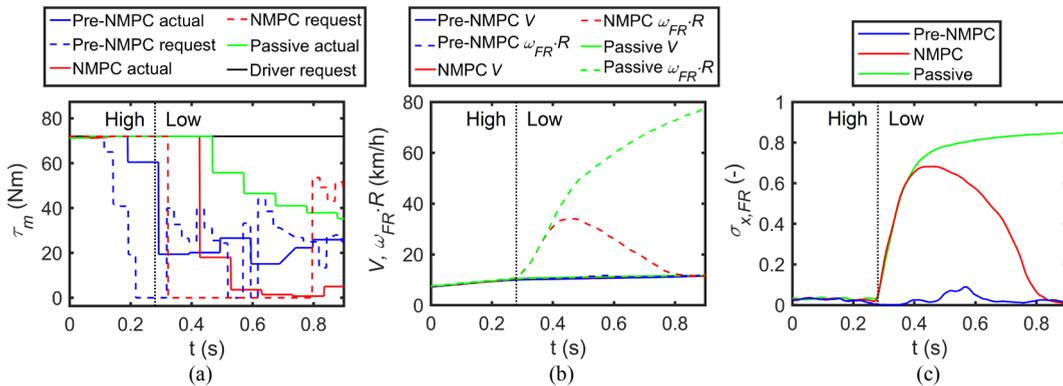

Fig. 3. Time profiles during experimental traction control tests: (a) driver torque request ('Driver request'), modified torque request ('NMPC request' and 'Pre-NMPC request') output by the traction controllers, and actual front motor torque ('Passive actual,' 'NMPC actual' and 'Pre-NMPC actual'); (b) longitudinal vehicle speed, $V$, and front right (FR) tangential wheel speed, $\omega_{FR}R$; and (c) FR tire slip ratio, $\sigma_{x,FR}$. The vertical dotted line separates the high and low friction sections of the test.

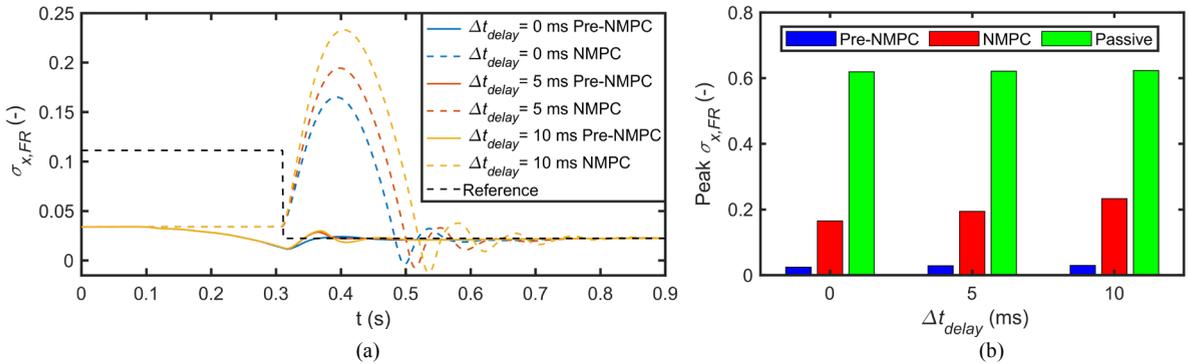

Fig. 4. Sensitivity analysis simulations for different values of the pure time delays of the electric powertrain: (a) time profiles of the FR slip ratio, $\sigma_{x,FR}$; and (b) peak values of $\sigma_{x,FR}$.

### 3.3. Pure time delay sensitivity simulations

A simulation-based sensitivity analysis during acceleration maneuvers with a sudden transition from high to low friction conditions was conducted to evaluate the performance of the NMPC traction controllers for different values of the pure time delay, $\Delta t_{delay}$, from the torque variation request (output by the traction controller) to the powertrain torque response. In the specific simulations, the controller setting is characterized by $H_P = 250$ ms and $N = 50$ steps, i.e., the long prediction horizon and high number of steps represent a best-case scenario, under the assumption of absence of control hardware limitations. In Fig. 4, for the Pre-NMPC algorithm the wheel slip ratio tracking performance is substantially independent of the pure time delay of the powertrain, thanks to the delay compensation algorithm in (11). In contrast, as expected, the benchmarking non-pre-emptive NMPC brings increasing slip ratio peaks for increasing values of $\Delta t_{delay}$, while the passive configuration shows consistently high wheel spinning.

## 4. Pre-emptive braking controller

### 4.1. Controller formulation

The double-track NMPC (DT-NMPC) formulation for pre-emptive braking uses a detailed (double-track) internal model, enabling advanced and predictive vehicle dynamics control, such as sideslip angle limitation, through the exclusive modulation of the traction/braking force at the vehicle level. This represents a novel cornering dynamics control method with minimum actuation complexity, as it does not involve the generation of any direct yaw moment.

The DT-NMPC prediction model equations are defined for a vehicle with a centralized on-board motor with a single powered axle, but can be adapted to any other configuration. The model is also expressed through the generic continuous time formulation in (1), with the state vector being:

$$\boldsymbol{x} = [S, V, \beta, \dot{\psi}, \omega_{FL}, \omega_{FR}, \omega_{RL}, \omega_{RR}]^T \tag{12}$$

where $S$ is the traveled distance; $\beta$ is the vehicle sideslip angle at the center of gravity; $\dot{\psi}$ is the yaw rate; and the subscripts '$Rj$' indicate the rear corners. The control input vector is:

$$\boldsymbol{u} = [\tau_{wh}, \varepsilon_V, \varepsilon_{\alpha_R}]^T \tag{13}$$

where $\tau_{wh}$ is the total wheel torque; and $\varepsilon_V$ and $\varepsilon_{\alpha_R}$ are the slack variables on vehicle speed and rear axle slip angle ($\alpha_R$), which impose soft constraints. For brevity, the 7-degree-of-freedom prediction model formulation is omitted from this paper, as it can be found in Guastadisegni et al. (2022).

The DT-NMPC shares the general NMPC cost function in (4), with the stage cost defined as:

$$l(\boldsymbol{x_n}, \boldsymbol{u_n}) = W_{u,\varepsilon_V}\varepsilon_{V,n}^2 + W_{u,\varepsilon_{\alpha_R}}\varepsilon_{\alpha_R,n}^2 + W_{u,\tau_{wh}}[\tau_{wh,driver} - \tau_{wh,n}]^2 \tag{14}$$

where $\tau_{wh,driver}$ is the driver wheel torque request; and $W_{u,\tau_{wh}}$, $W_{u,\varepsilon_V}$ and $W_{u,\varepsilon_{\alpha_R}}$ are the cost function weights for penalizing the torque deviation from the driver demand, and the slack variables.

The constraints are on the wheel torque, vehicle speed and rear axle slip angle:

$$\begin{aligned} \tau_{min} &\leq \tau_{wh,n} \leq \tau_{wh,driver} \\ V_n - \varepsilon_{V,n} &\leq V_{max,fut,n} \\ \varepsilon_{V,n} &\geq 0 \\ |\alpha_{R,n}| - \varepsilon_{\alpha_R,n} &\leq \alpha_{R,max} \\ \varepsilon_{\alpha_R,n} &\geq 0 \end{aligned} \tag{15}$$

The wheel torque $\tau_{wh,n}$ must remain within the limits, $\tau_{min}$ and $\tau_{wh,driver}$, related to the driver demand and available actuators, i.e., the friction brakes and electric powertrain, which are a hard constraint. $\alpha_{R,max}$ is the rear slip angle limit, which is set as a soft constraint, and is provided to the NMPC as an external parameter. Therefore, $\alpha_{R,max}$ remains constant along the prediction horizon, but can change during vehicle operation, e.g., as a function of the estimated tire-road friction condition. $V_{max,fut,n}$ is the speed limit at step $n$, calculated as:

$$V_{max,fut,n}(k) = \min\left(V_{veh,max}, V_{max,a_{y,max},n}(k)\right) = \min\left(V_{veh,max}, \sqrt{\frac{F_s\,\mu(k)\,g}{K_{ref,fut,n}(k)}}\right) \tag{16}$$

where $V_{veh,max}$ is the maximum vehicle speed in straight line conditions; $V_{max,a_{y,max},n}$ is the speed limit related to the road curvature $K_{ref,fut,n}$ and the estimated tire-road coefficient $\mu$ (which can be considered constant or variable along the prediction horizon); $F_s$ is a safety factor ($F_s \leq 1$); $g$ is the gravitational acceleration; and $K_{ref,fut,n}$ is the element $n$ of the future reference curvature vector, $\boldsymbol{K_{ref,fut}}$, which is obtained as a function of the vector of the expected future traveled distance values at the current time step $k$, $\boldsymbol{S_{fut}}(k)$:

$$\boldsymbol{K_{ref,fut}}(k) = f_{K_{ref,fut}}(\boldsymbol{S_{fut}}(k)) \tag{17}$$

where $f_{K_{ref,fut}}$ is set as a map, and $\boldsymbol{S_{fut}}(k)$ is calculated with (10).

### 4.2. Sideslip angle limitation through longitudinal vehicle dynamics control

A ground-breaking feature of the DT-NMPC algorithm is the capability of limiting sideslip angle to set levels, thanks to the road curvature preview, the high accuracy of its prediction model, the presence of the rear axle sideslip angle slack variable in (14), and the soft constraint in (15). This functionality is demonstrated through simulations of an obstacle avoidance test (see the ISO 3888 standard), which is frequently used to assess the performance of vehicle dynamics controllers. The objective is for the vehicle to complete the maneuver from high values of entry speed, without hitting any of the cones indicating the limits of the course.

Fig. 5 reports obstacle avoidance simulation results for the DT-NMPC with $\alpha_{R,max} = 3$, 6 and 9 deg. The absolute longitudinal coordinate $X$ is reported from -20 m, where $X = 0$ corresponds to the initial point of the course according to the ISO standard. The vehicle is set to be at the desired initial speed of 85 km/h at $X = -80$ m, after which the driver model tries to maintain constant speed, until the vehicle reaches $X = 0$, at which the driver model imposes zero wheel torque demand. The DT-NMPC makes braking interventions whenever necessary throughout the test.

In Fig. 5, as $\alpha_{R,max}$ decreases, the duration of the initial braking action is extended, which corresponds to lower maneuvering speeds. The different levels of vehicle speed translate into a variation of the sideslip angle dynamics consistent with the imposed constraints. The configuration with $\alpha_{R,max} = 9$ deg shows the largest maximum sideslip angle magnitude, which reaches almost -8 deg and experiences more oscillations than the other two configurations. Based on these results, the important novel conclusion is that in the next generation of stability controllers for connected vehicles, it will be possible to directly control sideslip angle by pre-emptively using only longitudinal vehicle dynamics control, which offers a new chassis control development route.

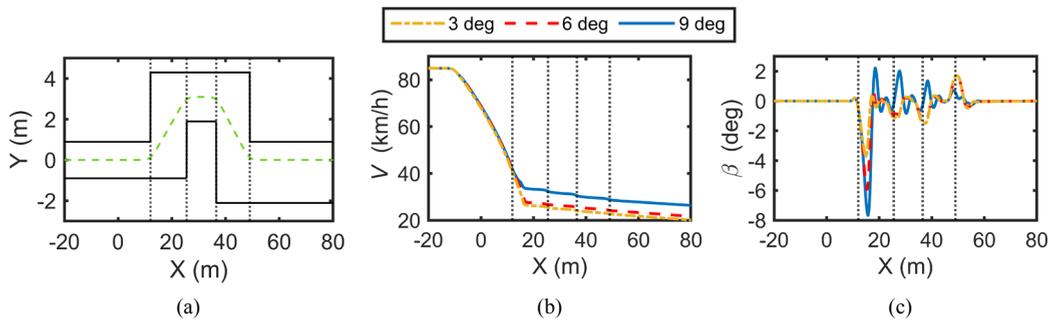

Fig. 5. Simulation results along obstacle avoidance tests from 85 km/h, carried out with the DT-NMPC with $\alpha_{R,max} = 3$, 6 and 9 deg: (a) reference path (dashed green line) and test track limits (solid lines) in the $X$ and $Y$ coordinates. The vertical dotted lines indicate the different sections of the course; (b) vehicle speed $V$ as a function of the $X$ coordinate; and (c) sideslip angle $\beta$ as a function of the $X$ coordinate.

*4.3. Real-time controller implementation and vehicle experiments*

As a proof-of-concept, the DT-NMPC algorithm was experimentally evaluated on the ZEBRA vehicle with $H_P = 3.4$ s and $N = 17$ steps, which is the limit condition for the real-time operation on the dSPACE platform of the ZEBRA EV. The EV includes an automated driver set-up, consisting of: i) a path tracking controller for generating the reference steering wheel angle, based on the sum of a feedforward contribution and a proportional integral (PI) feedback contribution using the lateral displacement and heading angle errors; and ii) a PI module, tracking the reference speed profile through the electric powertrain and friction brake actions. The reference path was programmed a priori, while the current vehicle position, speed and heading angle were identified through GPS and other on-board sensors.

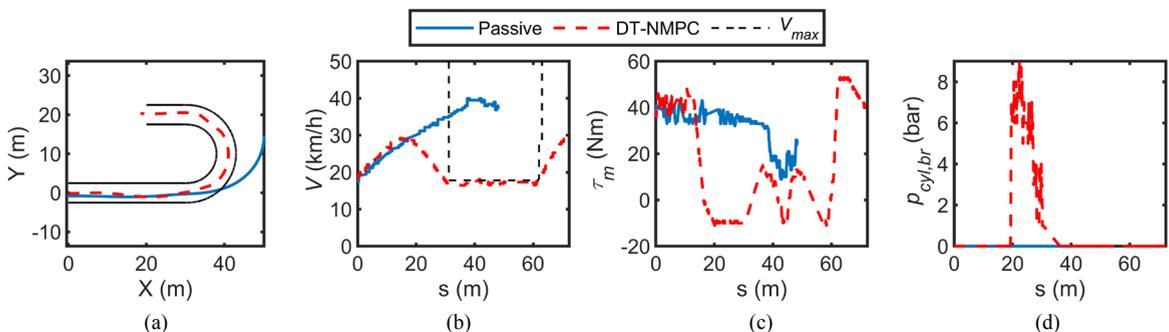

Fig. 6. Experimental results on the ZEBRA vehicle along a U-turn test: (a) trajectories in the $XY$ inertial frame; (b) vehicle speed, $V$; (c) motor torque, $\tau_m$; and (d) master cylinder pressure $p_{cyl,br}$, where the variables in (b)-(d) are expressed as functions of the distance $s$ along the reference path.

The results are reported in Fig. 6, for a U-turn maneuver, with straight line entry and exit sections, connected by a 10 m radius semi-circle. The speed at the entrance of the course is 20 km/h, while the target speed for the automated driver was set to 50 km/h. The passive vehicle is continuously accelerated throughout the test up to 40 km/h, and largely exceeds the speed at which the path tracking algorithm can keep the vehicle within the lane.

On the contrary, the DT-NMPC keeps the vehicle well within the lane boundaries, through the pre-emptive motor torque reductions and friction brake applications starting from $s = \sim 14$ m. The semi-circular trajectory section ends at $s = \sim 60$ m, at which the pre-emptive braking algorithm completes its intervention and the automated driver resumes control of the vehicle. The DT-NMPC vehicle speed complies with the $V_{max}$ profile from (16). This proof-of-concept experiment is an evident demonstration of the potential active safety benefits of the novel pre-emptive braking function.

## 5. Conclusion

This paper showcased the research on two pre-emptive vehicle controllers developed in recent and ongoing European projects, based on NMPC technology, using information from V2X connectivity. The first algorithm is a preview-based traction controller, which employs the information on the expected tire-road friction coefficient ahead to enhance the wheel slip control performance. The main highlights are:

- Proof-of-concept experiments on an EV demonstrator show that the traction controller with tire-road friction preview runs in real-time, and significantly enhances the wheel slip tracking performance, by decreasing the peak values of slip ratio by several times, during maneuvers characterized by sudden reductions of the tire-road friction level.
- Differently from the benchmarking reactive controller, the pre-emptive controller can compensate for the effect of the typical pure time delays of electric powertrains.

The second controller is a pre-emptive braking controller, i.e., the so-called DT-NMPC, for human-driven and automated vehicles, applying automated torque demand reductions if the current velocity level is projected to be excessively high with respect to the expected curvature profile ahead. The main conclusions are:

- Through road preview control, it is possible to enforce sideslip angle constraints during extreme obstacle avoidance tests without the application of any direct yaw moment. This opens up opportunities for the development of the next generation of vehicle dynamics controllers for connected vehicles.
- The DT-NMPC is real-time implementable, with prediction horizons exceeding 3 s as confirmed by the promising proof-of-concept experimental results on the ZEBRA vehicle.

For an in-depth discussion on the proposed controllers, readers may refer to publications by the same research team, namely Scamarcio et al. (2022) and Tavolo et al. (2022) for the pre-emptive traction controller, and Guastadisegni et al. (2022) for the pre-emptive braking controller. Future developments will focus on the implementation aspects of the pre-emptive controllers, including consideration of the required level of accuracy in vehicle localization.

## Acknowledgements

This work was supported by the European Union's Horizon 2020 program under grant agreement no. 769944 (STEVE), grant agreement no. 824254 (TELL), and grant agreement no. 101006953 (Multi-Moby).

## References

Guastadisegni, G., So, K.M., Parra, A., Tavernini, D., Montanaro, U., Gruber, P., Soria, L., Mantriota, G., Sorniotti A., 2022. Vehicle stability control through pre-emptive braking. International Journal of Automotive Technology (under review).
Houska, B., Ferreau, H.J., Diehl, M., 2011. An auto-generated real-time iteration algorithm for nonlinear MPC in the microsecond range. Automatica, vol. 47, no. 10, pp. 2279-2285.
Montanaro, U., Dixit, S., Fallah, S., Dianati, M., Stevens, A., Oxtoby, D., Mouzakitis, A., 2019. Towards connected autonomous driving: review of use cases. Vehicle System Dynamics, vol. 57, no. 6, p. 779–814.
Scamarcio, A., Caponio, C., Mihalkov, M., Georgiev, P., Ahmadi, J., So, K. M., Tavernini, D., Sorniotti, A., 2022. Predictive anti-jerk and traction control for V2X connected electric vehicles with central motor and open differential. IEEE Transactions on Vehicular Technology (in press).
Tavolo, G., So, K.M., Tavernini, D., Perlo, P., Sorniotti, A., 2022. Nonlinear Model Predictive Control for Preview-Based Traction Control, 15[th] International Symposium on Advanced Vehicle Control (AVEC). Kanagawa, Japan.
Zarkadis, K., Velenis, E., Siampis, E., Longo, S., 2018. Predictive Torque Vectoring Control with Active Trail Braking. European Control Conference. Limassol, Cyprus.